\documentclass{aa}  

\usepackage{amsmath} 

\usepackage{graphicx}
\usepackage{xcolor}
\usepackage{txfonts}
\usepackage{hyperref}
\usepackage{longtable}

\begin{document} 

\title{Investigating the origin of optical and X-ray pulsations of the transitional millisecond pulsar PSR J1023+0038}

\author{G.~Illiano\inst{1,2}, A.~Papitto \inst{1}, F.~Ambrosino\inst{1,3,4}, A.~Miraval Zanon\inst{1}, F.~Coti Zelati \inst{5,6,7}, L.~Stella\inst{1}, L.~Zampieri\inst{8}, A.~ Burtovoi\inst{9},
S.~Campana\inst{7}, P.~Casella\inst{1}, M.~Cecconi\inst{10},
D.~de Martino\inst{11}, M.~Fiori\inst{8}, A.~Ghedina\inst{10}, M.~Gonzales\inst{10}, M.~Hernandez Diaz\inst{10},  G.L.~Israel\inst{1}, F.~Leone\inst{13},
G.~Naletto\inst{12,8}, H.~Perez Ventura\inst{10}, C.~Riverol\inst{10}, L.~Riverol\inst{10}, D.F.~Torres\inst{5,6,14}, M.~Turchetta\inst{15}}

\institute{INAF-Osservatorio Astronomico di Roma, Via Frascati 33, I-00076, Monteporzio Catone (RM), Italy\\
\email{giulia.illiano@inaf.it}
\and Tor Vergata University of Rome, Via della Ricerca Scientifica 1, I-00133 Roma, Italy
\and INAF-Istituto di Astrofisica e Planetologia Spaziali, Via Fosso del Cavaliere 100, I-00133 Rome, Italy
\and Sapienza Università di Roma, Piazzale Aldo Moro 5, I-00185 Rome, Italy
\and Institute of Space Sciences (ICE, CSIC), Campus UAB, Carrer de Can Magrans s/n, E-08193, Barcelona, Spain
\and Institut d’Estudis Espacials de Catalunya (IEEC), Carrer Gran Capità 2–4, E-08034 Barcelona, Spain
\and INAF, Osservatorio Astronomico di Brera, Via E. Bianchi 46, I-23807 Merate (LC), Italy
\and INAF-Osservatorio Astronomico di Padova, Vicolo dell’Osservatorio 5, I-35122 Padova, Italy
\and INAF - Osservatorio Astrofisico di Arcetri, Largo Enrico Fermi 5, 50125, Florence, Italy
\and Fundación Galileo Galilei - INAF, La Palma, Spain
\and INAF–Osservatorio Astronomico di Capodimonte, Salita Moiariello 16, I-80131 Napoli, Italy
\and Department of Physics and Astronomy, University of Padova, Via F. Marzolo 8, 35131, Padova, Italy
\and Sezione Astrofisica, Dipartimento di Fisica e Astronomia, Università di Catania, Catania, Italy
\and Institució Catalana de Recerca i Estudis Avançats (ICREA), E-08010 Barcelona, Spain
\and Norwegian University of Science and Technology (NTNU), Høgskoleringen 1, 7491 Trondheim, Norway
}             

\date{}
\authorrunning{Illiano et al.}
\titlerunning{Optical and X-ray pulsations of the transitional millisecond pulsar PSR J1023+0038}

 
  \abstract
   {PSR J1023+0038 is the first millisecond pulsar that was ever observed as an optical and UV pulsar. So far, it is the only optical transitional millisecond pulsar. The rotation- and accretion-powered emission mechanisms hardly individually explain the observed characteristics of optical pulsations. A synergistic model, combining these standard emission processes, was proposed to explain the origin of the X-ray/UV/optical pulsations.}
   {We study the phase lag between the pulses in the optical and X-ray bands to gain insight into the physical mechanisms that cause it.}
   {We performed a detailed timing analysis of simultaneous or quasi-simultaneous observations in the X-ray band, acquired with the \textit{XMM-Newton} and \textit{NICER} satellites, and in the optical band, with the fast photometers SiFAP2 (mounted at the 3.6 m Telescopio Nazionale Galileo) and Aqueye+ (mounted at the 1.8 m Copernicus Telescope). We estimated the time lag of the optical pulsation with respect to that in the X-rays by modeling the folded pulse profiles with two harmonic components.}
   {Optical pulses lag the X-ray pulses by $\sim 150 \, \mathrm{\mu s}$ in observations acquired with instruments (\textit{NICER} and Aqueye+) whose absolute timing uncertainty is much smaller than the measured lag. We also show that the phase lag between optical and X-ray pulsations lies in a limited range of values, $\delta \phi \in (0-0.15)$, which is maintained over timescales of about five years. This indicates that both pulsations originate from the same region, and it supports the hypothesis of a common emission mechanism. Our results are interpreted in the \textit{\textup{shock-driven mini pulsar nebula}} scenario. This scenario suggests that optical and X-ray pulses are produced by synchrotron emission from the shock that formed within a few light cylinder radii away ($\sim 100 \, \mathrm{km}$) from the pulsar, where its striped wind encounters the accretion disk inflow.}
   {}


   \maketitle
%

\section{Introduction}
Transitional millisecond pulsars (tMSPs) are rapidly rotating ($P \lesssim 10 \, \mathrm{ms}$), weakly magnetized ($\approx 10^8 - 10^9 \, \mathrm{G}$) neutron stars (NSs) that have been observed to swing between distinct states within a few days that are likely powered by different physical mechanisms. These sources are part of binary systems with low-mass ($M \lesssim 1 \, M_\odot$) companion stars. At high-mass accretion rates, an accretion-powered pulsar that is able to channel the inflowing matter toward the NS magnetic poles is observed. When the accretion rate decreases, the pulsar wind sweeps the matter transferred through Roche-lobe overflow away, and a rotation-powered radio pulsar is seen. Variability in the mass inflow rate is the driver of these state changes.  
    
Three confirmed tMSPs are known to date: PSR J1023+0038 \citep{2009Sci...324.1411A}, XSS J1227-4859 \citep{2010A&A...515A..25D, 10.1093/mnras/stu708}, and IGR J1824-2452  \citep{Papitto_2013}. All of them have also been observed in an intermediate state, called subluminous disk state. The physical mechanism that powers this state still remains to be fully understood. In this state, tMSPs show the presence of an accretion disk and $\gamma$-ray emission that is up to ten times more intense than the emission that is observed during the rotation-powered state \citep[][and references therein]{Papitto_DeMartino_2020}. This is at variance with low-mass X-ray binaries (LMXBs), which generally do not emit a detectable $\gamma$-ray flux. Another peculiarity of tMSPs is the X-ray luminosity ($L_X \sim 10 ^{33} - 10^{34} \, \mathrm{erg \, s^{-1}}$), which is lower than what is usually observed in the outburst phase of accreting millisecond X-ray pulsars (AMXPs) ($L_X \sim 10^{36} \, \mathrm{erg \, s^{-1}}$), but higher than the luminosity in the rotation-powered state ($L_X <10^{32} \, \mathrm{erg \, s^{-1}}$). The X-ray emission is also variable over timescales of a few tens of seconds: different intensity modes (high, low, and flaring modes) have been observed in the X-ray light curves \citep[see, e.g.,][]{2013A&A...550A..89D, Linares_2014, Bogdanov_2015_2, archibald2015accretionpowered, Coti_Zelati_2018}.
J1023+0038 has been in the subluminous disk state from June 2013 \citep{Patruno_2014, Stappers_2014} until the time of writing (October 2022) and is in high mode for $\sim
80\%$ of the time \citep{Bogdanov_2015_2, archibald2015accretionpowered, Jaodand}. Transitions from high to low mode, and vice versa, occur unpredictably on a timescale of $\sim 10 \, \mathrm{s}$. The duration of these modes varies from a few tens of seconds to a few hours \citep{Papitto_DeMartino_2020}. 
    
J1023+0038 is the first millisecond pulsar that was ever observed as an optical pulsar \citep{Ambrosino_2017, 10.1093/mnrasl/slz043, Karpov_2019, Burtovoi_2020}. Optical pulsations have recently also been observed from the AMXP SAX J1808.4-3658 \citep{Ambrosino_2021}, but J1023+0038 so far remains the only tMSP with detectable pulsed emission in the optical band. Optical and X-ray pulsations from J1023+0038 were detected simultaneously in the X-ray high modes, but they disappeared when the source transited in the low modes. This suggests a common emission mechanism \citep{Papitto_2019}.
Recently, pulsations have been observed in the UV band. The UV emission, like the optical emission, undergoes transitions between high and low intensity modes that occur simultaneously with those observed in the X-ray band \citep{Jaodand_2021, Miraval_Zanon_2022}.
    
Previous works attempted to determine the physical origin of optical pulsations \citep[see, e.g.,][]{2019A&A...629L...8C, Papitto_2019, Veledina_2019}. Individually, the standard rotation- and accretion-powered mechanisms hardly explain the observed optical pulsed luminosity \citep{Ambrosino_2017, Papitto_2019}. X-ray pulsations were first interpreted as resulting from the channeling of matter along the magnetic field lines with the subsequent formation of accretion columns at the NS poles \citep{archibald2015accretionpowered}. However, the optical luminosity of accretion columns is expected to be much lower than the observed optical pulsed luminosity ($L_{\mathrm{opt}} \sim 10^{31} \, \mathrm{erg \, s^{-1}}$; \citealt{Ambrosino_2017}). 
Even assuming that optical pulses are generated by cyclotron emission from electrons falling into the accretion columns, the expected luminosity would be about 40 times lower than observed \citep{Ambrosino_2017, Papitto_2019}. Similar energetic arguments can also exclude the reprocessing of accretion-powered X-ray emission at the surface of the companion star and/or in the outermost regions of the disk, as occurs in some X-ray binaries with a strongly magnetized and slowly rotating accreting pulsar \citep{Ambrosino_2017}.

On the other hand, optical emission driven by the rotation of the NS magnetic field would require an efficiency in converting the spin-down power to the pulsed optical emission up to $10^4$ times higher than the values ($\sim 5 \times 10^{-6} - 2 \times 10^{- 9}$) measured for the five isolated rotation-powered pulsars from which optical pulses were detected \citep{Cocke_1969Nature, Mignani_2011, Ambrosino_2017}. Moreover, the fraction of spin-down power converted into X-ray pulses would be much higher than that of almost all rotation-powered pulsars \citep{Papitto_2019}. Consequently, the rotation-powered mechanism by itself cannot be the common origin of X-ray, UV, and optical pulsations from J1023+0038.
    
These implications led to the proposal of the shock-driven mini pulsar nebula scenario. Optical and X-ray pulsations originate from synchrotron emission in a shock that forms beyond the light cylinder radius, where the striped pulsar wind meets matter from the inner accretion disk \citep{Papitto_2019, Veledina_2019}. In this region, electrons are accelerated to relativistic speeds and emit synchrotron radiation by interacting with the magnetic field in the shock region. This configuration permits a higher fraction of the spin-down energy to be converted into X-ray pulses compared to the previously discussed cases. Pulsar wind nebulae indeed radiate up to a few percent of the pulsar spin-down power \citep{2008AIPC..983..171K, 2010AIPC.1248...25K, 2011ApJ...727..131V, Torres_2014}.
    
Analyses of simultaneous observations performed in May 2017 with \textit{XMM-Newton} and the fast optical photometer SiFAP2, mounted at the INAF Telescopio Nazionale Galileo (TNG), found that optical pulses lag the X-ray pulses by $\sim 200 \, \mathrm{\mu s}$ \citep{Papitto_2019}. The proposed model interprets this time lag in terms of the different timescales that synchrotron X-ray and optical photons take to be emitted.
However, the above measurement was affected by the absolute timing accuracy of SiFAP2 ($\sim 60 \, \mathrm{\mu s}$; \citealt{Papitto_2019}) and by that of  \textit{XMM-Newton}/EPIC.
In a Calibration Technical Note of May 2022\footnote{\url{https://xmmweb.esac.esa.int/docs/documents/CAL-TN-0220.pdf}}, the uncertainties on arrival times acquired through \textit{XMM-Newton}/EPIC were reviewed and reached a value of $100 \, \mathrm{\mu s}$ for the timing mode, which is more than twice that calculated by \citet{Martin_Carrillo_2012} of $48 \, \mathrm{\mu s}$ and considered by \citet{Papitto_2019} to estimate the significance of the optical/X-ray pulse phase lag they measured. The absolute timing accuracy of these instruments makes the systematic error associated with time lag estimated in \citet{Papitto_2019} compatible with the measure itself. This highlights the importance of presenting an in-depth study of the relation between optical and X-ray pulsations that also analyzes simultaneous observations acquired only with \textit{NICER} and the fast optical photometer Aqueye+, mounted at the Copernicus Telescope in Asiago. The \textit{NICER} absolute timing accuracy is estimated to be $< 300$ ns \footnote{\url{https://heasarc.gsfc.nasa.gov/docs/nicer/mission_guide/}}, while that of Aqueye+ is $< 0.5 \, \mathrm{ns}$ \citep{Zampieri_2015}. We report here a detailed timing analysis performed on optical/X-ray simultaneous or quasi-simultaneous observations to elucidate the physical mechanisms that cause the observed pulsations. Our data were acquired with the \textit{XMM-Newton} and \textit{NICER} X-ray satellites and with the fast optical photometers SiFAP2 and Aqueye+ over a time ranging from May 2017 to the beginning of February 2022. \\
   
Section \ref{sec:obs} is dedicated to the description of the observations and the data processing techniques. In Sect. \ref{sec:data_analysis}, we perform the phase analysis of simultaneous or quasi-simultaneous observations in the optical and the X-ray bands. By modeling the pulse profiles as the sum of two harmonic components, we study the time lags between the pulsations in the two different observational bands. We discuss our results in Sect. \ref{sec:discussion} and constrain the synergistic model proposed to explain the emission mechanisms of optical and X-ray pulsation from J1023+0038. Last, in Sect. \ref{sec:conclusions}, we summarize our main results and outline future prospects.\\
\section{Observations} \label{sec:obs}
Table \ref{table:1} lists the observations analyzed in this paper. They were selected with the aim of studying (quasi-)simultaneous observations of J1023+0038 in the optical and X-ray bands in a time interval of about five years. In the following, we detail the analysis of the different data sets.
    
\begin{table*}
\caption{Log of the (quasi-)simultaneous X-ray/optical observations of PSR J1023+0038.}             
\label{table:1}      
\centering                          
\begin{tabular}{l c c c}     
\hline\hline                 
 \textup{Telescope/Instrument (Obs. ID)} & \textup{Start Time (MJD)}$^a$ & \textup{Exposure (s)} & \textup{Band} \\   
 \hline
 \textit{2017 May}\\
TNG/SiFAP2 & 57896.9700580 & 3298 & white filter\\ 
TNG/SiFAP2 &  57897.8908020 & 8397 & white filter\\ 
\textit{XMM-Newton}/EPIC-pn (0794580801) &  57896.9293980 &  24914 & 0.3–10 keV\\
\textit{XMM-Newton}/EPIC-pn (0794580901) & 57897.7392740 & 23413 & 0.3–10 keV\\
\hline                        
\textit{2018 December 11 - 12}\\
Copernicus/Aqueye+ & 58464.0446059 & 1799 & white filter \\

Copernicus/Aqueye+ & 58464.0686652 & 1799 & white filter \\

Copernicus/Aqueye+ & 58464.0925580 & 2699 & white filter \\

Copernicus/Aqueye+ & 58464.1329206 & 2699 & white filter \\

Copernicus/Aqueye+ &  58464.1667650 & 2699 & white filter \\

Copernicus/Aqueye+ & 58464.2083977 & 1799 & white filter \\

\textit{XMM-Newton}/EPIC-pn  (0823750301) & 58463.8833467 & 30000 & 0.3–10 keV\\

\hline
\textit{2018 December 15}\\
Copernicus/Aqueye+  & 58467.0313365 & 1199 & white filter \\

Copernicus/Aqueye+ &  58467.0504218 & 3599 & white filter \\

Copernicus/Aqueye+  & 58467.0951940 & 1799 & white filter \\

Copernicus/Aqueye+  & 58467.1306531 & 1799 & white filter \\

Copernicus/Aqueye+  & 58467.1566119 & 1799 & white filter \\

Copernicus/Aqueye+ & 58467.1820657 & 1799 & white filter\\

Copernicus/Aqueye+  & 58467.2137354 & 1199 & white filter\\

\textit{XMM-Newton}/EPIC-pn (0823750401) & 58467.9157557 & 34000 & 0.3–10 keV\\

\hline
\textit{2019 January}\\
TNG/SiFAP2 &  58514.9781481 & 3300 & u filter$^b$ \\ 

\textit{NICER} (1034060118) & 58514.9150460 & 2268 & 0.2–12 keV\\
\textit{NICER} (1034060119) & 58514.9805560 & 6785  & 0.2–12 keV\\

\hline
\textit{2019 February}\\
Copernicus/Aqueye+  & 58520.0611407 & 4499 & white filter\\

Copernicus/Aqueye+  & 58520.8774590 & 3599 & white filter\\

Copernicus/Aqueye+  & 58520.9395999 & 3599 & white filter\\

Copernicus/Aqueye+ & 58521.0002527 & 3599 & white filter\\

\textit{NICER} (1034060120) & 58519.8725930 & 4265 & 0.2–12 keV\\
\textit{NICER} (1034060121) & 58520.0084100 & 3709 & 0.2–12 keV\\
\hline
\textit{2019 June}\\
TNG/SiFAP2 &  58636.8837037 &  2400 &  white filter\\                   

TNG/SiFAP2      & 58636.9135648 & 1200 & white filter\\                 

TNG/SiFAP2 & 58636.9194676 & 1560 & white filter\\      

\textit{NICER}  (2034060101) &  58636.8678240 &  1185 & 0.2–12 keV\\

\hline 
\textit{2020 January}\\
Copernicus/Aqueye+ &  58879.0814000 & 1799 & white filter\\ 

Copernicus/Aqueye+ &  58879.1064600 & 1799 & white filter \\

TNG/SiFAP2 & 58878.9843981 & 3600 & white filter \\ 

TNG/SiFAP2 & 58879.0284954 &  3600 &  white filter\\ 

TNG/SiFAP2 &   58879.0722454 &  3600 &  white filter \\ 

TNG/SiFAP2 &   58879.1163426 &   3600 &  white filter \\ 

\textit{NICER}  (2034060110) & 58878.9268060 & 2265 & 0.2–12 keV\\
\textit{NICER} (2034060111) &  58878.9921300 &  11533 & 0.2–12 keV\\

\hline 
\textit{2022 January-February}\\
Copernicus/Aqueye+ &  59608.0382707 & 13395 & white filter\\ 
Copernicus/Aqueye+ &  59609.0113612 & 13750 & white filter\\
Copernicus/Aqueye+ &  59609.9358742 & 13465 & white filter\\
Copernicus/Aqueye+ &  59611.9862628 & 5955 & white filter\\
Copernicus/Aqueye+ &  59613.0046037 & 11870 & white filter\\
\textit{NICER}  (4034060110) & 59607.9890106 & 4771 & 0.2–12 keV\\
\textit{NICER}  (4034060111) & 59609.0265844 & 2496 & 0.2–12 keV\\
\textit{NICER}  (4034060112) & 59609.9946040 & 2910 & 0.2–12 keV\\
\textit{NICER}  (4034060113) & 59611.9905482 & 1480 & 0.2–12 keV\\
\textit{NICER}  (4034060114) & 59613.0439361 & 617 & 0.2–12 keV\\

\hline           
\\
\multicolumn{4}{l}{
\begin{minipage} {12.7cm}
\tiny $^a$: Barycentric dynamical time at exposure start.\\
$^b$: SDSS u filter with $\lambda_{eff}= 349 \, \mathrm{nm}$ and $\Delta \lambda_{FWHM}=68 \, \mathrm{nm}$. This observation was part of a campaign in which u, g, and r filters were used to investigate any dependences of the pulse amplitude on the spectral band. 
\end{minipage}}
\end{tabular}
\end{table*}

\subsection{X-ray observations}

\subsubsection{\textit{NICER}}
We present the analysis of \textit{NICER} observations of J1023+0038 that were simultaneously or quasi-simultaneously performed with optical observations from January 2019 to February 2022. The events were reduced and processed using \texttt{HEASoft} version 6.28 and \texttt{NICERDAS} version 7a.
We corrected the photon arrival times to the Solar System Barycenter (SSB) using the JPL ephemerides DE405. We adopted the source coordinates R.A. (J2000)=10:23:47.687198(2) and DEC. (J2000)=$+$00:38:40.84551(4) \citep{2012ApJ...756L..25D}. We use these coordinates throughout the rest of the work.    
We estimated the background contributions to our data with the tool \texttt{nibackgen3C50}  \citep{2022AJ....163..130R}.\\
Generally, it was sufficient to analyze the individual observations listed in Table \ref{table:1} to derive an orbital solution (Table \ref{table:2}), except in the case of the June 2019 data set, for which the close-by observation of 2019 June 13 (Obs. ID: 2531010401) had to be used as well.

\subsubsection{\textit{XMM-Newton}/EPIC}
The \textit{XMM-Newton} observations were performed on 2017 May 23 and 24 (presented in \citealt{Papitto_2019}), and on 2018 December 11 and 15 (Table \ref{table:1}). The data were reduced using the \texttt{Science Analysis Software} (\texttt{SAS}) v.16.1.0. In each observation, the EPIC-pn was operating with a time resolution of $29.5 \, \mathrm{\mu s}$ (timing mode) and a thin optical blocking filter. The photon arrival times observed by \textit{XMM-Newton} were reported to the SSB, using the JPL ephemerides DE405 and the \texttt{barycen} tool from \texttt{HEASoft}. We defined source and background regions with coordinates RAWX=27–47 and RAWX=3–5, respectively, and retained good events characterized by a single or a double pattern.

\subsection{Optical observations}
\subsubsection{TNG/SiFAP2}
We analyzed the optical observations reported in Table \ref{table:1} acquired with the SiFAP2 fast optical photometer \citep{Meddi_2012, doi:10.1142/S2251171716500057, Ambrosino_2017} mounted at the TNG.
The arrival times of each photon were referred to the SSB through the \texttt{TEMPO2} package \citep{tempo2}, using the JPL ephemerides DE405.\\
In the January 2019 data set, the statistics was too poor to obtain precise measurements of the epoch of passage at the ascending node, $T_{asc}$, and of the spin period, $P$ (Table \ref{table:2}). In this part of the analysis, we merged that observation with SiFAP2 observations carried out on 2019 February 1, with a total exposure of $\sim 10$ ks.\\
The SiFAP2 quartz clock is characterized by drifts with respect to the actual time measured by two global positioning system (GPS) pulse-per-second (PPS) signals that are used to mark the beginning and end of each observation. Following \citet{Ambrosino_2017}, the arrival times recorded by SiFAP2, $t_{\mathrm{SiFAP2}}$, were corrected assuming the linear relation $t_{\mathrm{arr}} = t_{\mathrm{SiFAP2}} \times (\Delta t_{\mathrm{GPS}} / \Delta t_{\mathrm{SiFAP2}})$, where  $\Delta t_{\mathrm{GPS}}$ and $\Delta t_{\mathrm{SiFAP2}}$ are the total elapsed time measured by the GPS and SiFAP2 clocks, respectively.
In January 2019, we had $\Delta t_{\mathrm{SiFAP2}} - \Delta t_{\mathrm{GPS}} = - 1.164 \, \mathrm{ms}$, with the value of $\Delta t_{\mathrm{GPS}}$ reported in Table \ref{table:1}. During the observations of June 2019, we had $\Delta t_{\mathrm{SiFAP2}} - \Delta t_{\mathrm{GPS}} = - 2.323 , -1.156 \, \mathrm{, and} -1.511  \, \mathrm{ms}$. Finally, for January 2020 observations, we had $\Delta t_{\mathrm{SiFAP2}} - \Delta t_{\mathrm{GPS}} = +0.580 , +0.751, +0.822 \, \mathrm{, and} +0.663  \, \mathrm{ms}$.

\subsubsection{Copernicus/Aqueye+}
Aqueye+ is an ultra-fast optical single photon counter mounted at the Asiago 1.8-meter Copernicus Telescope with the capability of time-tagging the detected photons with subnanosecond time accuracy \citep{Zampieri_2015}.
The chosen Aqueye+ observations (Table \ref{table:1}) were reduced with the \texttt{QUEST} software (v. 1.1.5, see \citealt{Zampieri_2015}). The arrival times of each photon were referred to the SSB through the \texttt{TEMPO2} package, using the JPL ephemerides DE405. 

\section{Data analysis} \label{sec:data_analysis}
For each data set, we first corrected the photon arrival times for the pulsar orbital motion in the binary system. We set the orbital period and the projected semimajor axis equal to the values found in the timing solution of \citet{Jaodand} (see Table \ref{table:2}), and we performed a search on the epoch of passage at the ascending node, $T_{asc}$. We used a grid of $T_{asc}$ values spaced by 0.37 s \citep{2012MNRAS.427.2251C} around the estimate extrapolated from \citet{Jaodand}. We carried out an epoch-folding search on each time series by sampling each spin period, $P$, with 16 phase bins. The final best $T_{asc}$ was determined by fitting the peak of the $\chi^2$ distribution with a Gaussian function. We considered the half width at half maximum (HWHM) of the Gaussian as the uncertainty to associate with $T_{asc}$. To improve the spin period estimate obtained from the $T_{asc}$ search, we corrected the photon arrival times with the best-fitting values of the orbital parameters and performed an epoch-folding search with 16 phase bins. The best $P$ value was then estimated by modeling the peak of the $\chi^2$ distribution with a Gaussian function. The associated error was calculated using Eq. (6a) from \citet{1987A&A...180..275L}. Table \ref{table:2} summarizes the best-fitting values of $T_{asc}$ and $P$ found for each data set.
\begin{table*}
\caption{Summary of the orbital parameter estimates for different observations of PSR J1023+0038.}             
\label{table:2}      
\centering                          
\begin{tabular}{l l l l }     
\hline\hline                 
Fixed parameters & & &  \\
\hline
Orbital period (d), $P_{orb}$ & 0.1980963155$^a$\\
Projected semimajor axis (lt-s), $x \equiv a_1 \sin{i}$ & 0.343356$^a$\\
R.A. (J2000), $\alpha$ & 10:23:47.687198$^b$\\
DEC. (J2000), $\delta$ & $+$00:38:40.84551$^b$\\
Position epoch (MJD) & 55000$^b$\\
Orbital eccentricity, $e$ & 0 \\
\hline 
Telescope/Instrument & $P$ (ms) & $T_{asc}$ (MJD) & $T_{start}$ (MJD) \\
\hline
\textit{2017 May}\\
\textit{XMM-Newton}/EPIC & $1.6879874455(24)$ & $57896.829263(13)$ & 57896.92939840\\
TNG/SiFAP2 & $1.6879874449(58)$ & $57896.829275(38)$ & 57896.97005780\\
\hline
\textit{2018 December 11 - 12}\\
\textit{XMM-Newton}/EPIC & $1.6879874440(48)$ & $58463.781107(14)$ & 58463.88334670\\
Copernicus/Aqueye+ & $1.687987425(27)$ & $58463.781111(27)$ & 58464.04460940\\
\hline 
\textit{2018 December 15}\\
\textit{XMM-Newton}/EPIC &  $1.6879874445(42)$ & $58467.743034(14)$ & 58467.91575570\\
Copernicus/Aqueye+ & $1.687987439(25)$ & $58467.743029(22)$ & 58467.03133470\\
\hline
\textit{2019 January}\\
\textit{NICER} & $1.687987447(19)$ & $58514.889969(18)$ & 58514.92190000 \\
TNG/SiFAP2 &  $1.687987533(48)$ & $58515.08803(17)$ & 58514.98403960\\
\hline
\textit{2019 February}\\
\textit{NICER} & $1.687987442(11)$ & $58519.8423788(99)$ & 58519.87965120\\
Copernicus/Aqueye+ & $1.687987389(60)$ & $58519.842377(28)$ & 58519.87965120\\
\hline
\textit{2019 June} \\
\textit{NICER} & $1.687987446(96)$ & $58636.9173512(24)$ & 58636.86910000 \\
TNG/SiFAP2 &  $1.687987446(70)$ & $58636.91736(18)$ & 58636.88410956\\
\hline 
\textit{2020 January} \\
\textit{NICER} & $1.687987446(13)$ & $58878.991096(21)$ & 58878.93440000 \\
TNG/SiFAP2 &  $1.6879874844(69)$ & $58878.9911009(12)$ & 58878.99024860\\
Copernicus/Aqueye+ & $1.6879874451(24)$ & $58878.991104(11)$ & 58877.98782000\\
\hline 
\textit{2022 January-February} \\
\textit{NICER} & $1.6879874474(22)$ & $59607.985869(18)$ & 59607.98901060 \\
Copernicus/Aqueye+ &  $1.6879874574(85)$   & 59607.985862(33) & 59607.98901060\\

\hline  
\tiny$^a$: From \citet{Jaodand}.\\
\tiny$^b$: From \citet{2012ApJ...756L..25D}.
\end{tabular}
\end{table*}
We verified that the timing results were compatible between simultaneous X-ray and optical observations.
This allowed us to use the values of $T_{asc}$ and $P$ found from the X-ray timing (which are more accurate than those obtained from the analysis of optical observations due to the higher root mean square (rms) pulse amplitude of the signal, i.e., the pulse amplitude divided by the square root of 2) to correct the photon arrival times for the pulsar orbital motion and to perform the phase analysis of simultaneous optical and X-ray observations. Since the folded pulse profiles are double-peaked \citep[see, e.g.,][]{archibald2015accretionpowered, Ambrosino_2017, Papitto_2019}, we modeled them using a decomposition function with two harmonic terms,
\begin{equation} \label{eq:two_harms}
    F(\phi) = K \, \Biggr\{ 1 + \sum_{i=1}^2 \, r_i \, \sin{[2 \, \pi \, i \, (\phi - \phi_i)]} \Biggr\}, 
\end{equation}
where $K$ is the average count rate, and the free parameters, $r_i$ and $\phi_i$ ($i= 1, 2$), are the fractional amplitude and the phase of the two harmonics, respectively. Uncertainties of our best-fitting values were estimated from the parameter range required to increase the $\chi^2$ from the fit of a quantity $\Delta \chi^2(\alpha = 68 \%) = 1.0$ \citep{Lampton_Margon_Bowyer, 1976ApJ...210..642A, Yaqoob_1998}. 

\subsection{Evolution of the time of passage at the ascending node}
\begin{figure}
   \centering
   \includegraphics[width=9.1cm]{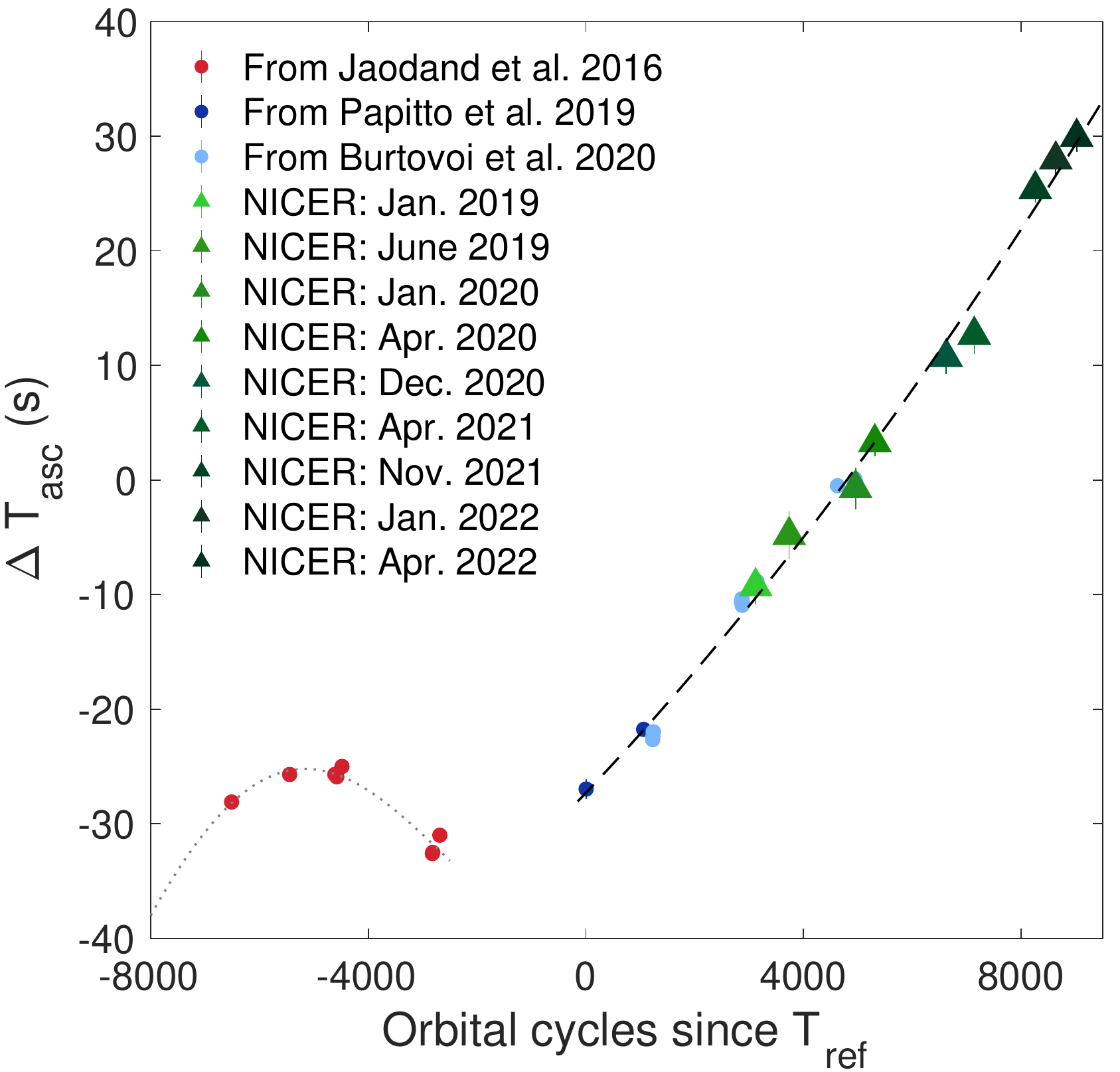}
   \caption{Long-term evolution of $T_{asc}$ as a function of the number of orbital cycles since $T_{ref}=57897.027668 \, \mathrm{MJD}$. Red points are the values found by \citet{Jaodand}, which deviate from the almost linearly increasing trend found in this work. Blue points are from \citet{Papitto_2019}, and lighter blue points are from \citet{Burtovoi_2020}. Triangles, associated in the legend with the different \textit{NICER} observations, are from this work. The thick dashed line indicates our increasing roughly linear trend. The dotted gray line indicates the sinusoidal trend that is similar to the trend found in previous work \citep[e.g.,][]{Jaodand, Papitto_2019}.} \label{Fig:Tasc}
   \end{figure}

Figure \ref{Fig:Tasc} shows the difference $\Delta T_{asc}$ between our values of $T_{asc}$ in \textit{NICER} observations (Table \ref{table:2}) and that computed using the radio timing solution \citep{archibald2013longterm, Jaodand} as a function of the number of orbital cycles since $T_{ref}=57897.027668 \, \mathrm{MJD}$. A similar approach was adopted in previous works \citep[see, e.g.,][]{Jaodand, Papitto_2019, Burtovoi_2020}. \textit{NICER} observations in addition to those chosen in this paper simultaneously with optical data (Table \ref{table:1}) were analyzed to study the $T_{asc}$ long-term evolution (Table \ref{table:Tasc_additional}). We selected observations with an exposure $> 10 \, \mathrm{ks}$ in order to have good statistics for the timing analysis and at least two measurements of $T_{asc}$ per year.
\begin{table} \small
\renewcommand{\arraystretch}{1.2}
\centering
\caption{Additional observations of PSR J1023+0038 that were used to study the evolution of the epoch of passage at the ascending node.}              
\label{table:Tasc_additional}      
\begin{tabular}{l l l}          
\hline\hline                        
Telescope (Obs. ID) & $T_{asc}$ (MJD) & $T_{start}$ (MJD)\\
\hline
\textit{NICER} (3515010101) & 58949.117238(14) & 58949.1224730\\
\textit{NICER} (3515010802) & 59209.019690(17) & 59209.0096634\\
\textit{NICER} (4531010203) & 59311.039315(19)& 59311.0110863 \\
\textit{NICER} (4531010601) & 59533.303528(13) & 59533.3494883 \\
\hline
\end{tabular}
\end{table}
\noindent
\noindent
We found an increasing trend of $\Delta T_{asc}$ with time, as already inferred since May 2017 by \citet{Burtovoi_2020}, who emphasized that the steady increase may indicate a systematic underestimation of the orbital period of the system. We modeled the second part of the data in Fig. \ref{Fig:Tasc}, that is, from  $T_{ref}=57897.027668 \, \mathrm{MJD}$, with the following expression:
\begin{equation}
\Delta T_{asc}(N_{orb}) = A + B \, N_{orb} + \frac{1}{2} \, P_{orb} \, C \, N_{orb}^2.
\end{equation}
$P_{orb}$ is the orbital period, which was fixed at the value given in Table \ref{table:2}, while $A$, $B$, and $C$ are free parameters. The integer number of orbital cycles since $T_{ref}$ is $N_{orb} = \textup{int} [(T-T_{ref})/P_{orb}]$. We obtained $A = (-27.3 \pm 0.8)$ s, $B = (5.0 \pm 0.4) \times 10^{-3}$ s, and $C = (1.7 \pm 0.5) \times 10^{-11} \, \mathrm{s/s}$.
In general, redbacks, that is, millisecond pulsars in a close orbit with a low-mass companion star such as the source under consideration, show unpredictable variations of the orbital phase \citep[see, e.g.,][Fig. 2]{Jaodand}. 
Therefore, we caution that the $B$ term might be related to the correction we should apply to the orbital period to have constant residuals over time and $C$ as the orbital period derivative.
Since the orbital period difference would not significantly affect the results of the analysis presented here, we retained the value reported by \citet{Jaodand}. Future studies will allow us to firmly establish the binary evolution.

\subsection{Nonselection of intensity modes} \label{sec:modes}
\begin{figure*}
    \centering
    \includegraphics[width=19cm]{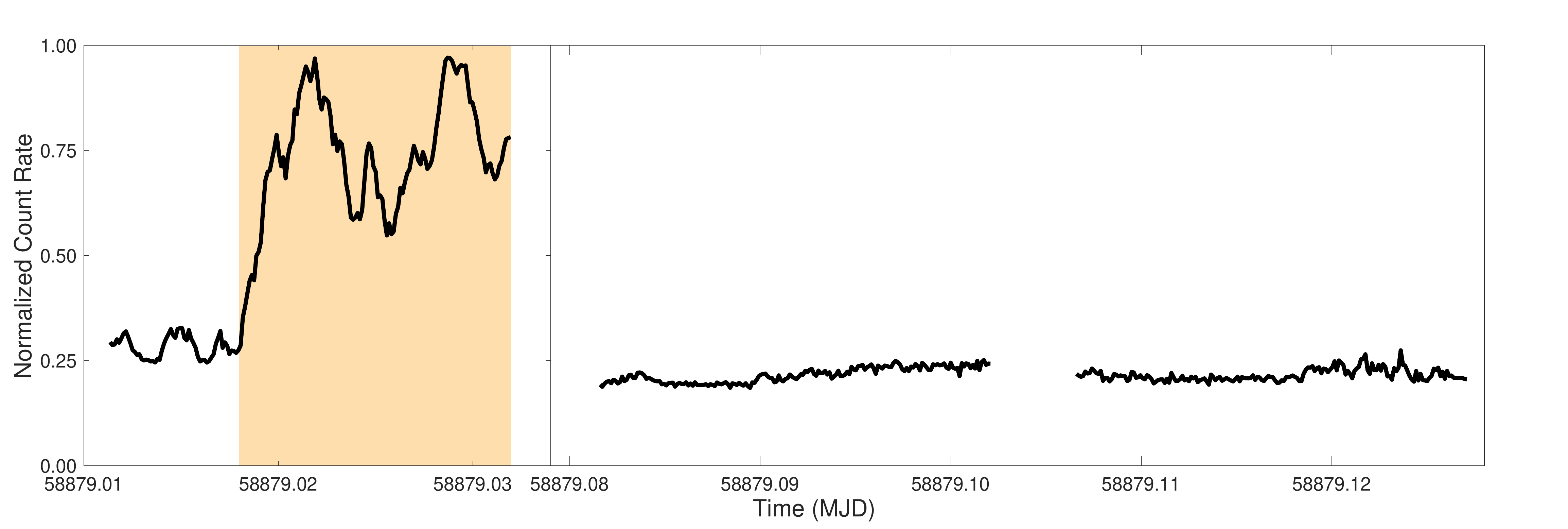}
    \centering
    \caption{Aqueye+ light curve observed in January 2020, binned every $10 \, \mathrm{s}$. The count rate is normalized at the maximum count rate. The vertical colored region indicates the visually identified flaring interval.}\label{fig:LC}
\end{figure*}

Except for the May 2017 data set, we did not distinguish among low, high, and flaring modes, differently from what was done in other works \citep[see, e.g.,][]{Bogdanov_2015_2, Papitto_2019}. As we show in the following, the nonselection of modes does not change the phase of our signal. The reason for this choice lies in the fact that different modes are typically identified in the X-ray band: just a fraction of the optical data considered in this work is strictly simultaneous to X-ray observations, differently from the case of the 2017 observations analyzed in \citet{Papitto_2019} (hereafter P19). It is not possible to precisely distinguish high and low modes in the optical band because the corresponding variations in intensity are fainter than those observed in the X-rays and are preferentially observed in the red part of the visible spectrum \citep[see, e.g.,][]{Shahbaz_2015, Shahbaz_2018, Kennedy_2018, Papitto_2018}.\\
To verify whether the pulse phase changes depend on either selecting or not selecting the different modes, we analyzed the optical observation acquired in May 2017 (Table \ref{table:1}) in three different ways: by selecting only the high modes; by using the whole observation, therefore without distinguishing between low, high, and flaring modes; and by visually selecting and removing flares, hence keeping both low and high modes. We first selected high modes only in simultaneous intervals with the X-ray observation, adopting the definition of X-ray modes from \citet{Bogdanov_2015_2}. Because our values of $T_{asc}$ and $P$ (Table \ref{table:2}) are compatible with those used by P19, equal to $P_{ref} = 1.687987446019$ ms and $T_{asc}=57896.82926(1)$ MJD, we used the latter values for a better comparison. The first and second harmonics of the optical pulsation lag the X-ray harmonics by $\delta \tau_1 = (223 \pm 31) \, \mathrm{\mu s}$ and $\delta \tau_2 = (195 \pm 7) \, \mathrm{\mu s}$ (Table \ref{table:dt}), respectively. Our results are compatible with those of P19. We note that the phases of the optical pulse profile obtained by selecting only the high modes are $\phi_1  = 0.4726 \pm 0.0053$ and $\phi_2 = 0.5500 \pm 0.0023$ (Table \ref{table:phi}).
Second, analyzing the entire optical observations of May 2017 without selecting the different intensity modes, we obtained $\phi_1  = 0.4714 \pm 0.0070$ and $\phi_2 = 0.5615 \pm 0.0034$. These differ by $0.2\sigma$ and $5\sigma$, respectively, from the phases estimated when only high-mode intervals were selected.
When we removed the flaring intervals (i.e., analyzing both high and low modes), the results are compatible with those obtained by selecting only high modes.
In this third case, we indeed obtained $\phi_1  = 0.4666 \pm 0.0064$ and $\phi_2 = 0.5501 \pm 0.0029$, both compatible within $3\sigma$ with the results from the data set in which only the high modes were analyzed. This is expected because optical (and X-ray) pulses are not detected during the low modes, whereas optical pulsations appear during flares, although the pulse amplitude is six times smaller than in the high modes \citep{Papitto_2019}.
For this reason, we paid special attention to removing flaring intervals in the remainder of this work, but we did not distinguish between high and low modes. Optical flares are indeed easy to identify \citep[see, e.g.,][]{Bogdanov_2015_2, Papitto_2018, Papitto_2019}. As an example, Fig. \ref{fig:LC} shows the visual selection of flares in the Aqueye+ observation of January 2020. We conclude that, with these caveats, the decision not to distinguish high and low modes does not produce different results in the phase analysis.

\subsection{Phase analysis results}
We summarize our main results in Tables \ref{table:phi} and \ref{table:dt}. We filtered X-ray and optical observations to analyze simultaneous intervals. The cases in which this was not possible are marked in the tables with an asterisk. The reason was either the absence of exact simultaneity or the short duration of these intervals, which made the statistics insufficient to detect the pulsed profile. Table \ref{table:phi} lists the best values for the phases and amplitudes, $\phi_i$ and $r_i$, with $i=1,2$, obtained by modeling the pulse profiles with a function consisting of two harmonics (Eq. \eqref{eq:two_harms}). $T_{ref}$ denotes the reference epoch (the same for simultaneous optical/X-ray observations) against which we calculated the phases. Fractional amplitudes were converted to background-subtracted rms amplitudes, that is, $r' = (1/\sqrt{2}) \, r \, c_{\mathrm{tot}}/(c_{\mathrm{tot}}-c_{\mathrm{bkg}}),$ where $r$ is the pulse amplitude, $c_{\mathrm{tot}}$ is the total count rate, and $c_{\mathrm{bkg}}$ is the count rate associated with the sky background. $R=(r_1^2+r_2^2)^{1/2}$ is the total rms amplitude. We associated the statistical errors computed through the least-squares method, setting  $\Delta \chi^2 (\alpha = 68 \%) = 1.0$ (Sect. \ref{sec:data_analysis}). The absolute timing accuracies for \textit{XMM-Newton} and SiFAP2 are $\sim 100 \, \mathrm{\mu s}$ and $\sim 60 \, \mathrm{\mu s}$, respectively, while systematic errors are negligible for \textit{NICER} and Aqueye+. Table \ref{table:dt} shows the corresponding lags between optical and X-ray pulsations.

\begin{figure*}
    \centering
    \includegraphics[width=15cm]{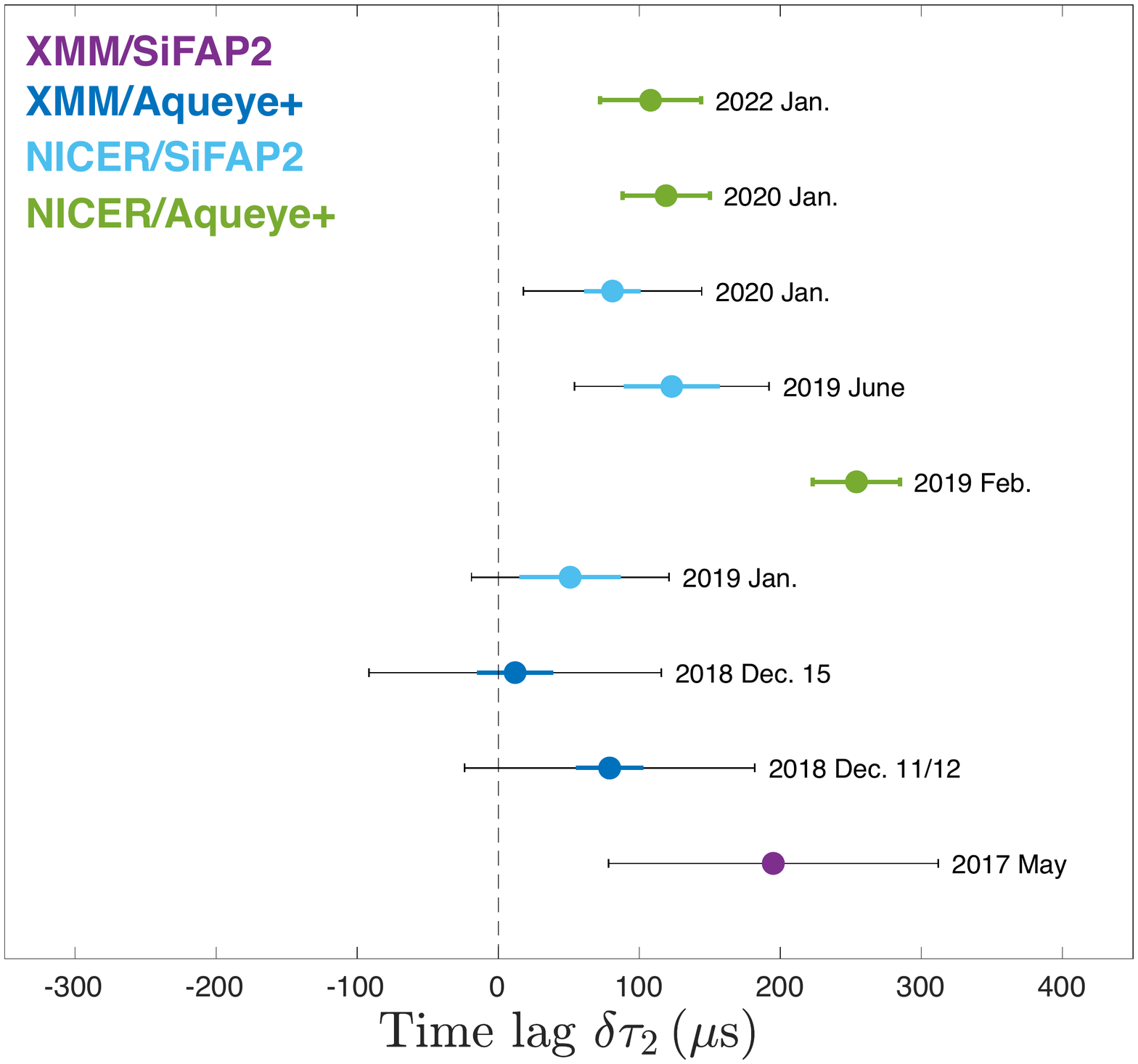}
    \caption{Time lags relative to the second harmonic term (Table \ref{table:dt}). The y-axis is not temporally equispaced. The purple point indicates the values for simultaneous observations between \textit{XMM-Newton} and SiFAP2, blue points show values between \textit{XMM-Newton} and Aqueye+, light blue points indicate values between \textit{NICER} and SiFAP2, and green points show the values between \textit{NICER} and Aqueye+. The dashed line indicates a zero time lag. Colored error bars represent 1$\sigma$ statistical uncertainties, while the black error bars indicate the total error. The influence of the absolute timing accuracy of \textit{XMM-Newton} ($\sim 100 \, \mathrm{\mu s}$) and SiFAP2 ($\sim 60 \, \mathrm{\mu s}$) is visible.} \label{Fig:dt2}
\end{figure*}

\begin{table*}
\caption{Properties of the X-ray and optical pulses.}             
\label{table:phi}      
\centering                          
\begin{tabular}{l c c c c c c c}     
\hline\hline                 
Telescope/Instrument & $T_{ref}$ (MJD) & $\phi_1$  & $\phi_2$ & $r_1 \, (\%)$ & $r_2 \, (\%)$ & $R \, (\%)$\\   
\hline                        
\textit{2017 May} - overlap: $11.0 \, \mathrm{ks}$\\
\textit{XMM-Newton}/EPIC  & 57896.0 & $0.340(18)$ & $0.4346(39)$ & 3.34(39) & 7.69(38) & 8.38(38) \\
TNG/SiFAP2 & 57896.0 & $0.4726(53)$ & $0.5500(23)$ & 0.391(13) & 0.449(13) & 0.595(13)\\
\hline
\textit{2018 December, 11/12} - overlap: $10.8 \, \mathrm{ks}$\\ 
\textit{XMM-Newton}/EPIC & 58463.0 & $0.474(24)$ & $0.1177(42)$ & 2.01(32) & 5.86(31) & 6.29(31) \\
Copernicus/Aqueye+ & 58463.0 & $0.564(29)$ & $0.164(14)$ & $0.145(48)$ & $0.159(48)$ & 0.215(48)\\
\hline 
\textit{2018 December, 15} - temporal gap: $41 \, \mathrm{ks}$*\\ 
\textit{XMM-Newton}/EPIC & 58467.0 & $0.256(34)$ & $0.4271(43)$ & 1.30(28) & 5.11(28) & 5.27(28)\\
Copernicus/Aqueye+ & 58467.0 & $0.315(23)$ & $0.434(16)$ & $0.167(39)$ & $0.128(39)$ & 0.210(39) \\
\hline 
\textit{2019 January} - overlap: $2.3 \, \mathrm{ks}$\\ 
\textit{NICER} & 58514.0 & $0.048(34)$ &  $0.1665(78)$ & $2.78(61)$ & $6.12(60)$ & 6.72(60) \\
TNG/SiFAP2 & 58514.0 & $0.153(98)$ & $0.197(20)$ & 0.30(17) & 0.70(17) & 0.76(17)\\
\hline 
\textit{2019 February} - overlap: $1.1 \, \mathrm{ks}$\\ 
\textit{NICER}  & 58519.0 & $0.288(26)$ &  $0.3831(76)$ & $3.01(50)$ & $5.12(49)$ & 5.94(49)\\
Copernicus/Aqueye+ & 58519.0 & $0.385(32)$ & $0.534(16)$ & $0.122(60)$ & $0.118(59)$ & 0.170(60)\\
\hline 
\textit{2019 June} - overlap: $340 \, \mathrm{s}$*\\
\textit{NICER} & 58636.0 & $0.560(87)$ &  $0.353(16)$ & $2.6(1.3)$ & $6.8(1.3)$ & 7.3(1.3)\\
TNG/SiFAP2 & 58636.0 & $0.775(42)$ & $0.426(13)$ & 0.085(22) & 0.139(22) & 0.163(22) \\
\hline 
\textit{2020 January} - overlap: $4.6 \, \mathrm{ks}$\\ 
\textit{NICER}  & 58878.0 & $0.171(56)$ &  $0.234(12)$ & $2.77(98)$ & $6.53(96)$ & 7.09(96) \\
TNG/SiFAP2  & 58878.0 & $0.1917(55)$ & $0.2823(23)$ & 0.2651(95) & 0.3286(95) & 0.4222(95) \\
\hline 
\textit{2020 January} - overlap: $520 \, \mathrm{s}$\\
\textit{NICER}  & 58878.0 & $0.207(65)$ &  $0.221(12)$ & $3.9(1.5)$ & $10.1(1.6)$  & 10.8(1.6) \\
Copernicus/Aqueye+  & 58878.0 & $0.204(30)$ & $0.292(14)$ & 0.240(57) & 0.256(57) & 0.351(57) \\
\hline 
\textit{2022 January} - overlap: $1.7 \, \mathrm{ks}$\\ 
\textit{NICER}  & 59607.0 & $0.058(75)$ &  $0.717(10)$ & $1.19(55)$ & $4.13(54)$ & 4.30(54) \\
Copernicus/Aqueye+  & 59607.0 & $0.031(82)$ & $0.781(19)$ & 0.064(57) & 0.135(57) & 0.149(57) \\
\hline     
\\
\multicolumn{7}{l}{
\begin{minipage} {18cm}
\tiny $^*$: Cases in which it was not possible to analyze exactly simultaneous intervals between optical and X-ray observations.
\end{minipage}
  }
\end{tabular}
\end{table*}
\begin{table*}
\caption{Lags between optical and X-ray pulsations.}             
\label{table:dt}      
\centering                          
\begin{tabular}{l c c c c c c }     
\hline\hline                 
 Date & Opt. Instrument & X-ray Telescope & $\delta \phi_1$ & $\delta \tau_1 \, (\mathrm{\mu s})  $  & $\delta \phi_2 $ & $\delta \tau_2 \, (\mathrm{\mu s})  $\\     
\hline                        
\textit{2017 May} & SiFAP2 & \textit{XMM-Newton} & $1.32(18) \times 10^{-1}$ & $2.23(31) \times 10^{2}$ & $1.154(45) \times 10^{-1}$ & $1.950(70) \times 10^{2}$\\

\textit{2018 December, 11/12} & Aqueye+ & \textit{XMM-Newton} & $9.0(3.8) \times 10^{-2}$ & $1.52(64) \times 10^{2}$ & $4.7(1.4) \times 10^{-2}$ & $7.9(2.4) \times 10$\\ 

\textit{2018 December, 15}* & Aqueye+ & \textit{XMM-Newton} & $5.9(4.1) \times 10^{-2}$ & $9.9(7.0) \times 10$ & $0.7(1.6) \times 10^{-2}$ & $1.2(2.7) \times 10$\\

\textit{2019 January} & SiFAP2 & \textit{NICER} & $1.0(1.0) \times 10^{-1}$ & $1.8(1.7) \times 10^{2}$ & $3.0(2.1) \times 10^{-2}$ & $5.1(3.6) \times 10$\\ 

\textit{2019 February} & Aqueye+ & \textit{NICER} & $9.7(4.1) \times 10^{-2}$ & $1.63(69) \times 10^{2}$ & $1.50(18) \times 10^{-1}$ & $2.54(31) \times 10^{2}$\\ 

\textit{2019 June}* & SiFAP2 & \textit{NICER} & $2.15(97) \times 10^{-1}$ & $3.6(1.6) \times 10^{2}$ & $7.3(2.0) \times 10^{-2}$ & $1.23(34) \times 10^{2}$\\ 

\textit{2020 January} & SiFAP2 & \textit{NICER} & $2.1(5.6) \times 10^{-2}$ & $3.5(9.5) \times 10$ & $4.8(1.2) \times 10^{-2}$ & $8.1(2.0) \times 10$\\ 

\textit{2020 January} & Aqueye+ & \textit{NICER} & $-0.4(7.1) \times 10^{-2}$ & $-0.1(1.2) \times 10^{2}$ & $7.0(1.8) \times 10^{-2}$ & $1.19(31) \times 10^{2}$\\

\textit{2022 Junuary} & Aqueye+ & \textit{NICER} & $-0.3(1.1) \times 10^{-1}$ & $-0.5(1.9) \times 10^{2}$ & $6.4(2.1) \times 10^{-2}$ & $1.08(36) \times 10^{2}$\\

\hline     
\\
\multicolumn{7}{l}{
\begin{minipage} {14cm}
\tiny $^*$: Cases in which it was not possible to analyze exactly simultaneous intervals between optical and X-ray observations.\end{minipage}}
\end{tabular}
\end{table*}
Figure \ref{Fig:dt2} focuses on the results from the second harmonic, whose power spectral densities are more than three times higher than those of the first harmonic. Colored error bars represent 1$\sigma$ statistical uncertainties, and the black error bars indicate the total error. We note the influence of the absolute timing accuracy of \textit{XMM-Newton} ($\sim 100 \, \mathrm{\mu s}$) and SiFAP2 ($\sim 60 \, \mathrm{\mu s}$). The time lags of optical and X-ray pulsations are always within the range $(0-250) \, \mathrm{\mu s}$, that is, a phase lag of $\delta \phi \in (0-0.15)$, even in observations acquired over five years. 

\subsubsection{\textit{NICER} and Aqueye+ simultaneous observations} \label{sec:discussion_NICER_AQ}
In this section, we discuss the results of (quasi-)simultaneous observations between Aqueye+ and \textit{NICER}. They are most valuable for our analysis because the absolute timing uncertainties are negligible compared with the effects we aim to measure. They were also taken with a different set of instruments than in P19, thus with disconnected systematics and of a much smaller magnitude. We have simultaneous observations between these two instruments in February 2019, January 2020, and January-February 2022 (Table \ref{table:1}).

\begin{figure}
   \centering
   \includegraphics[width=9cm]{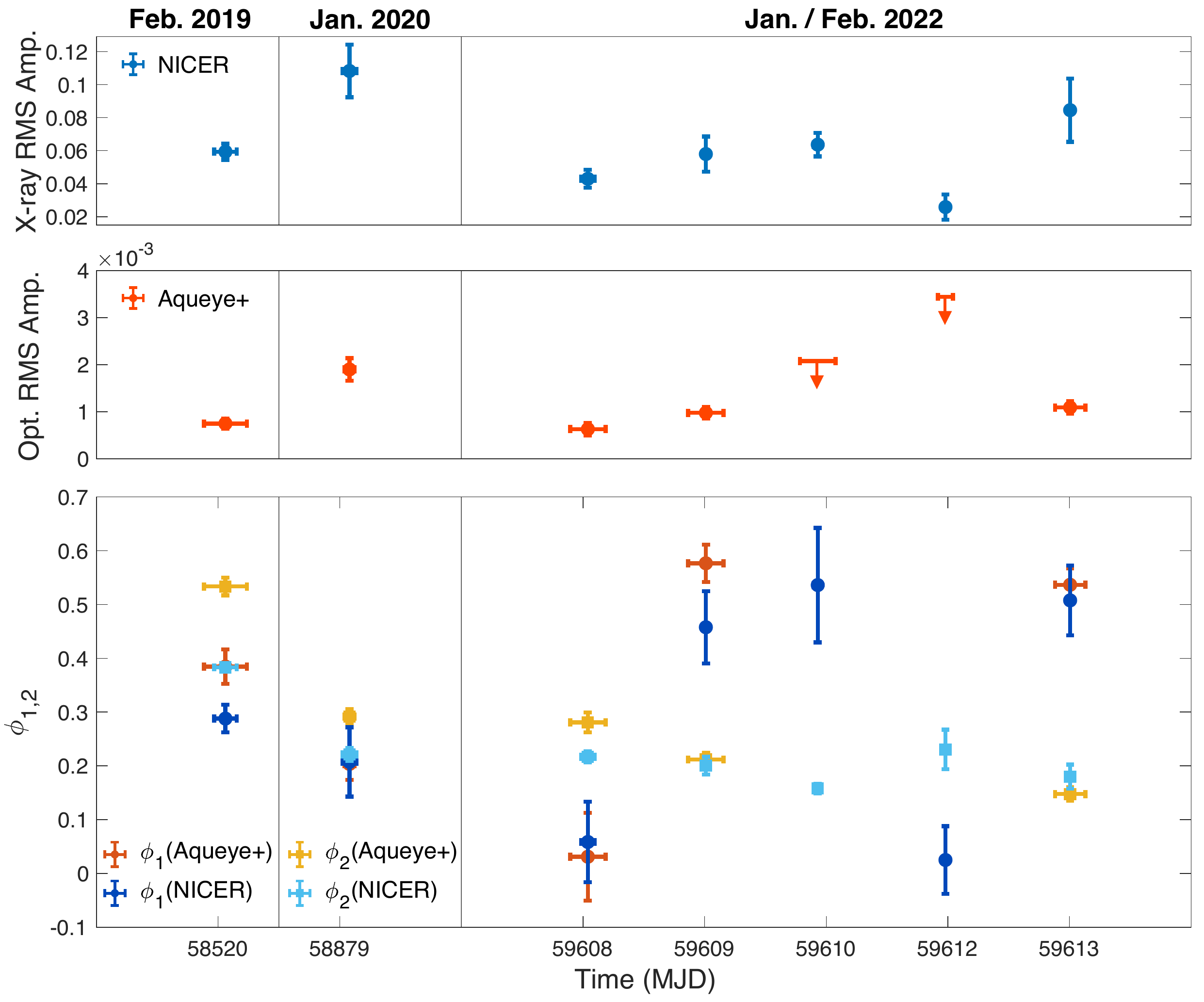}
      \caption{Background-subtracted rms amplitudes of X-ray (top panel) and optical (middle panel) pulse during \textit{NICER}/Aqueye+ simultaneous observations in February 2019, January 2020, and January-February 2022. The arrows represent the upper limits converted into rms amplitudes at $3 \sigma$ confidence level computed when the optical pulsations were not detected. In the lower panel, we show the phases of the optical pulse (red dots and empty yellow squares show the first and second harmonic, respectively), and the phases of the X-ray pulse (blue dots and light blue empty squares show the first and second harmonic, respectively) during these simultaneous observations.}
      \label{Fig:plot_JanFeb2022}
\end{figure}

From January 29 to February 2, 2022 an observational campaign was carried out with Aqueye+ and \textit{NICER}. However, optical data were affected by the bad weather conditions, and the X-ray statistics was often low due to short \textit{NICER} exposure. The top and middle panels of Fig. \ref{Fig:plot_JanFeb2022} show the X-ray and optical rms amplitudes in February 2019, in January 2020, and over the five-day observational campaign in 2022. When optical pulsations were not detected, we estimated upper limits on the pulse amplitude by computing the Fourier power spectral density and measuring the power of the first and second harmonics of the spin frequency. We then converted these power pairs into rms amplitudes at $3 \sigma$ confidence level according to the procedure described by \citet{Vaughan_1994} to take into account that the probability distribution of total power in a frequency bin of a Fourier spectrum containing both signal and noise is more complicated than a $\chi^2$ distribution. The bottom panel shows the phase of the two harmonic components of the corresponding optical and X-ray pulse profiles. In the main analysis of this work (see Tables \ref{table:phi} and \ref{table:dt}, and Fig. \ref{Fig:dt2}), we report the results from the first day of the 2022 observational campaign (January 29), when we had the longest X-ray observation that was suitable to provide statistically acceptable results and the weather conditions were good enough to detect optical pulsations. The phase of the second harmonic term of optical pulses lags that in the X-rays by $\delta \tau _2 = (108 \pm 36) \, \mathrm{\mu s}$.

Special attention was paid to filtering out flare intervals present in the optical data of February 2019 and January 2020. We obtained time lags from the second harmonic component of $\delta \tau _2 = (254 \pm 47) \, \mathrm{\mu s}$ and $(119 \pm 31) \, \mathrm{\mu s}$, respectively (Table \ref{table:dt}).

\begin{figure}
   \centering
   \includegraphics[width=9cm]{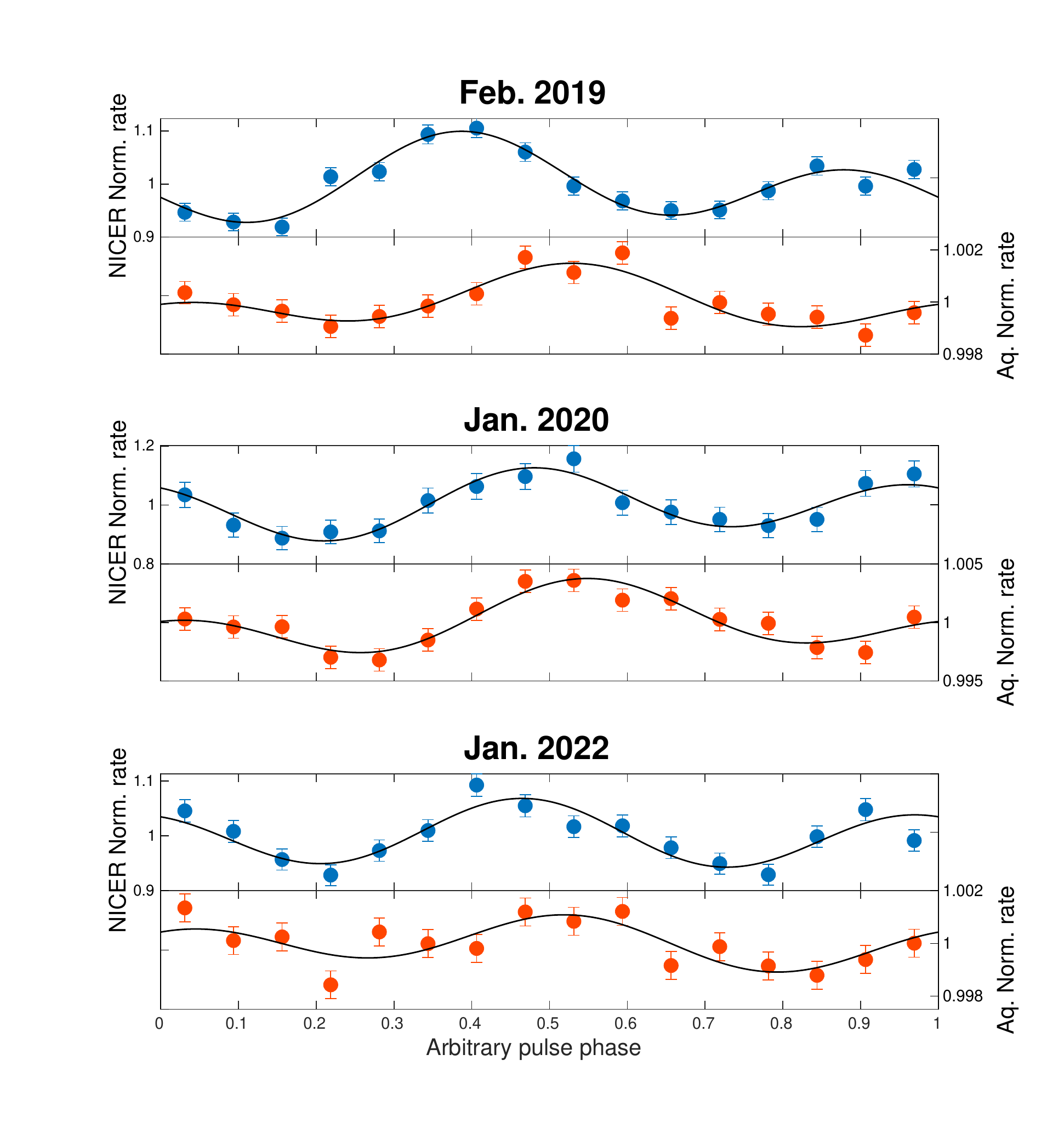}
          \caption{Pulse profiles obtained by folding with 16 phase bins of simultaneous observations with \textit{NICER} (blue points, left y-axis scale) and Aqueye+ (red points, right y-axis scale) in February 2019 (top panel), January 2020 (middle panel), and January 2022 (bottom panel). Solid lines indicate the best-fitting two-harmonic functions. The zero of the pulse phase was shifted arbitrarily in each panel by the same amount for both instruments.}
      \label{Fig:pulse_profiles}
\end{figure}

Figure \ref{Fig:pulse_profiles} shows the pulsed profiles of the \textit{NICER}/Aqueye+ simultaneous observations discussed here.

\begin{figure}
   \centering
   \includegraphics[width=9.2cm]{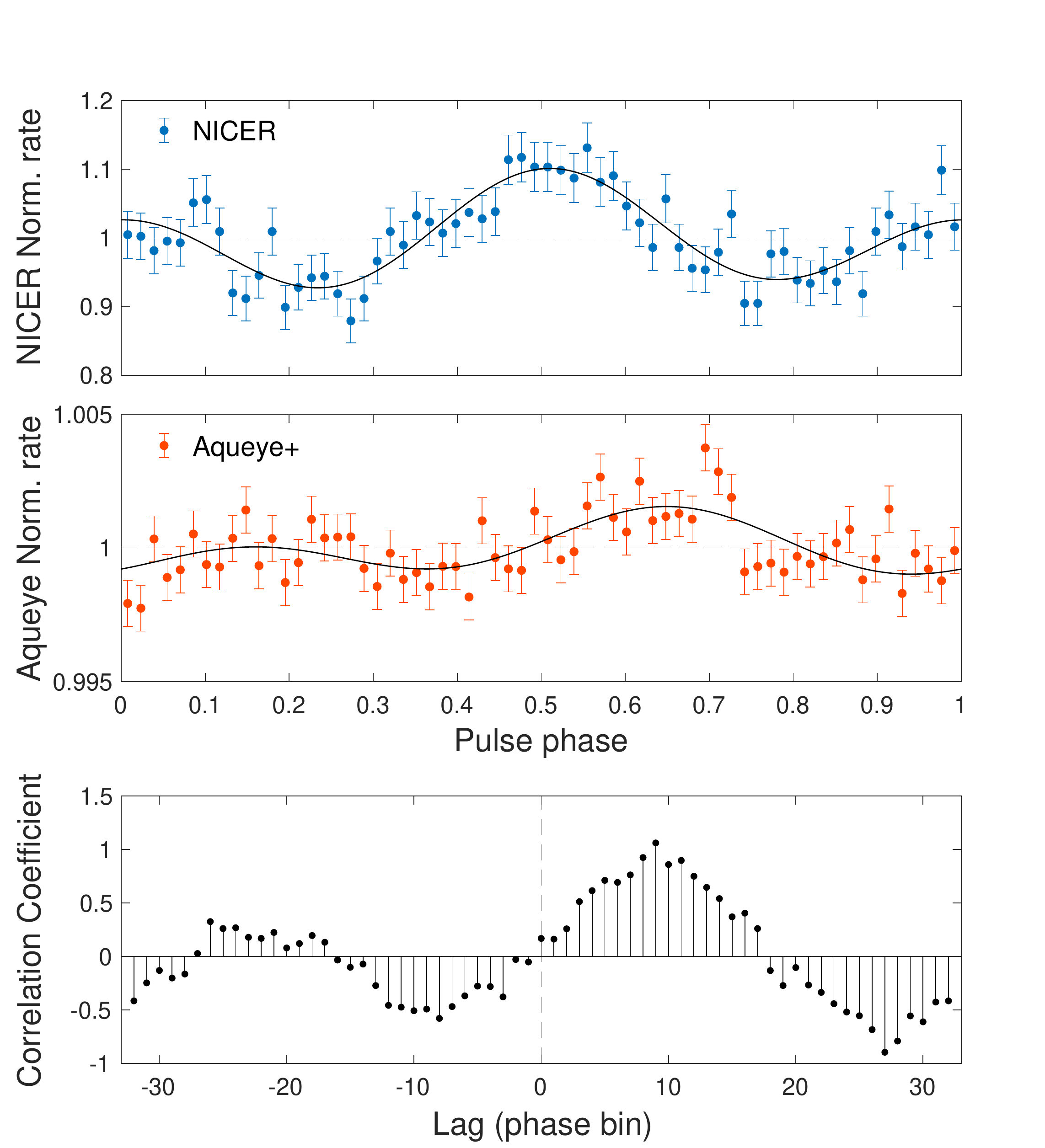}
      \caption{Cross-correlation of the pulse profiles from \textit{NICER} and Aqueye+ observations of February 2019 as the lag varies. The top and middle panels show \textit{NICER} and Aqueye+ pulse profiles with 64 phase bins, respectively. Solid lines indicate the best-fitting two-harmonic function. The bottom panel shows the cross-correlation coefficient as a function of the lag with which the optical pulse profile is shifted. We used phase lags ranging from $ - 32$ to $ + 32$, including the zero lag (dashed vertical line).
      }
      \label{Fig:cross}
   \end{figure}

Among the three sets of \textit{NICER}/Aqueye+ (quasi-) simultaneous observations discussed in this section, the February 2019 data feature the longest interval of simultaneity between \textit{NICER} and Aqueye+ (Table \ref{table:phi}), combined with the highest significance of the pulse profiles. Therefore, they can be considered most valuable for estimating the optical/X-ray time lag. This data set is also of particular interest because it shows the largest time lag with respect to the second harmonic term (Fig. \ref{Fig:dt2}). To strengthen the time lag measure obtained from this data set, we performed a cross-correlation analysis of the two pulse profiles as a function of the optical profile offset against that in X-rays.
The cross-correlation function \citep{cross_corr_function} was corrected for the counting statistics, and the data were wrapped around themselves to avoid leaving unmatched points. For instance, when we shifted the optical pulse profile of one phase bin to the right, the number of counts in the last bin of the optical profile was matched with the number of counts of the first bin of the X-ray profile. 
We folded the Aqueye+ and \textit{NICER} pulse profiles of February 2019 with a 64 phase bins.
Figure \ref{Fig:cross} shows that the maximum value of the correlation coefficient is at a lag of 9 phase bins. When 64 phase bins are used, each of them has a length equal to $P/64 \sim 26 \, \mathrm{\mu s}$, where $P$ is the spin period from the X-ray timing in $\mu s$ (Table \ref{table:2}). Therefore, this phase lag corresponds to a time lag of $\delta \tau \sim (237 \pm 13) \, \mathrm{\mu s}$, where the associated error is half the bin width. This result is significantly different from zero at more than 5$\sigma$ and is fully compatible with the analysis result for the second harmonic phase. This conclusion also shows that the second component prevails in the computation of the correlation between the optical and X-ray pulse profiles.

\subsubsection{Other observations} \label{sec:discussion_2}
We report the results of the remaining data sets, including observations carried out with \textit{XMM-Newton} and/or SiFAP2, which must be taken with caution as they are affected by larger absolute timing uncertainties than those of \textit{NICER} and Aqueye+.
Simultaneous or quasi-simultaneous observations performed between \textit{XMM-Newton} and Aqueye+ in December 2018 and between \textit{NICER} and SiFAP2 in January 2019 and in January 2020 return time lags that are compatible with the absence of a phase shift considering the systematic errors.
The observational campaign of June 2019 performed with SiFAP2 and \textit{NICER} provides $\delta \tau _2 =(124 \pm 34) \, \mathrm{\mu s}$, considering the statistical errors alone. When the SiFAP2 absolute timing accuracy is added in quadrature, we obtain $\delta \tau _2 =(130 \pm 69) \, \mathrm{\mu s}$.
An improvement in the absolute timing accuracy of \textit{XMM-Newton} and SiFAP2 would lead to statistically stronger results. 
Although the systematic errors affecting these estimates often make the lags compatible with zero, these observations are important to confirm that the time lag relative to the second harmonic term between optical and X-ray pulses is always in the $\sim(0 - 250) \, \mathrm{\mu s}$ range, that is, is not randomly distributed.

\section{Discussion} \label{sec:discussion}
This paper presented a detailed timing analysis of optical/X-ray (quasi-)simultaneous observations of the tMSP PSR J1023+0038, focusing on the study of the time lags between the pulses in the optical and X-ray bands.
We folded the data at the spin periods found in the X-ray timing, compatible with the values estimated in the simultaneous optical observations (Table \ref{table:2}), and we modeled the pulse profiles with two harmonic terms. \\
The optical pulses have total rms pulsed amplitudes of $\sim 0.1 - 0.8\%$, while the X-ray total rms pulsed amplitudes are in the range $4.3 - 10.8 \%$ (Table \ref{table:phi}), in agreement with what was found in previous works \citep{archibald2015accretionpowered, Bogdanov_2015_2, Ambrosino_2017, Papitto_2019, 10.1093/mnrasl/slz043, Miraval_Zanon_2022}.
Moreover, the rms pulsed fraction is variable over time (e.g., top and middle panels of Fig. \ref{Fig:plot_JanFeb2022}), as was also found in \citet{Miraval_Zanon_2022}.\\
Although the estimated time lags (Table \ref{table:dt}) are not consistent with being modeled by a single value, we report the weighted averages obtained from the studies of the first and second harmonic terms, respectively: $\overline{\delta \tau}_1 =(175 \pm 22) \, \mathrm{\mu s}$ and $\overline{\delta \tau}_2 =(162 \pm 6) \, \mathrm{\mu s}$. These results are compatible with each other within $1\sigma$.\\
By focusing on the second harmonic of the pulse profiles (because their power spectral densities are higher than those of the first harmonic), we found that the time lag between optical and X-ray pulsations always lies in the limited range of $(0-250) \, \mathrm{\mu s}$, also taking into account observations acquired over about five years (Fig. \ref{Fig:dt2}). We note the large total errors due to the absolute timing accuracy of \textit{XMM-Newton} ($\sim 100 \, \mathrm{\mu s}$) and SiFAP2 ($\sim 60 \, \mathrm{\mu s}$), highlighting the value of the results from \textit{NICER}/Aqueye+ simultaneous observations.

A timescale of $\sim 250 \, \mathrm{\mu s}$ is virtually equal to the light cylinder radius light-crossing time (for a NS spin frequency of $\nu \simeq 592.42$ Hz, $R_{LC} \simeq 80$ km), indicating that the optical pulsed luminosity might be produced by the reprocessing of accretion-powered X-ray emission. However, the observed pulsed luminosity in the visible band is too high to be explained in this way \citep{Ambrosino_2017, Papitto_2019}. The properties of optical and X-ray pulsations, such as their simultaneous detection during the X-ray high modes and their disappearance in the low modes \citep{Papitto_2019}, the similar pulse shape, and the very limited range of the estimated time lags might indicate that they are related, either by originating in the same region or in a region that lies very near, or that they are connected to the same emission process.

X-ray pulsations observed in the high modes of J1023+0038 were at first interpreted as due to the accretion of matter channeled along the magnetic field lines of the pulsar \citep{archibald2015accretionpowered}. However, this model fails to explain the observed optical pulsed luminosity of $\approx 10^{31} \, \mathrm{erg \, s^{-1}}$. When we assume that optical pulsations result from cyclotron emission by electrons that are in-falling in the accretion columns on the NS hotspots, the luminosity produced in this way is about 40 times lower than the observed one \citep{Ambrosino_2017, Papitto_2019}. 
 
On the other hand, indicating the efficiency in converting the spin-down power into the optical pulsed luminosity as $\eta _{\mathrm{opt}} = L_{\mathrm{opt}}/\dot{E}_{\mathrm{sd}}$, where $\dot{E}_{\mathrm{sd}}$ is the pulsar spin-down power, the measured values for the five rotation-powered pulsars from which optical pulses were detected \citep{Cocke_1969Nature, Mignani_2011} are in the range from $\eta _{\mathrm{opt}} \sim 5 \times 10^{-6}$ down to $\sim 2 \times 10^{- 9}$ (see red points in Fig. 3 from \citealt{Ambrosino_2017}). These NSs are all young ($10^3 - 10^5$ years), isolated, and possess high-magnetic fields ($> 10^{12}$ G). A higher value, equal to $\eta _{\mathrm{opt}} \sim 2 \times 10^{- 5}$, is found for J1023+0038. In addition, the fraction of the spin-down power converted into X-ray pulses would be much higher than that of almost all rotation-powered pulsars \citep{Papitto_2019}. The rotation-powered mechanism can hardly be the only common source of X-ray and optical pulsations from J1023+0038.

A model was thus proposed in which, despite the presence of an accretion disk, a rotation-powered pulsar is active in the system. In the shock-driven mini pulsar nebula scenario \citep{Papitto_2019, Veledina_2019}, optical and X-ray pulses are produced via synchrotron emission from a shock that is formed where the pulsar wind meets the accretion disk, within a few light cylinder radii from the pulsar. Inside the light cylinder, the magnetic field is described in terms of a split-monopole \citep{1999A&A...349.1017B}. The two monopoles with opposite sign give rise to a current sheet on the equatorial plane, which expands as an Archimedean spiral until it reaches the shock in two opposite spots where electrons are accelerated to relativistic speeds. Indicating the shock distance from the pulsar as $ \sim k\, R_{LC}$, for low values $k \simeq 1 - 2$, just beyond the light cylinder radius, the magnetic field is still so intense that synchrotron emission provides the dominant cooling mechanism for shock-accelerated electrons. 
This model is compatible with the presence of two harmonics. For quite large inclination angles, the emission from the spot on the observer's side is obscured by the disk. In contrast, the emission from the farthest spot, which is modulated sinusoidally during its rotation, is more easily observed (see Fig. 14 of \citealt{Papitto_2019}). The higher intensity of the second compared to the first harmonic may derive from an asymmetry of the system that causes them to be obscured in a different way.

We used a parametric value of the time lag between optical and X-ray pulsations to estimate some physical quantities of the system within this model. Based on our most accurate results, that is, those from \textit{NICER}/Aqueye+ simultaneous observations (Sect. \ref{sec:discussion_NICER_AQ}), we adopted $\delta \tau _p \sim 150 \, \mathrm{\mu s}$. 
Synchrotron photons are emitted by the shock-accelerated electrons on a timescale (Eq. (7) of \citealt{Papitto_2019})
\begin{equation} \label{eq:tsync_orig}
\begin{split}
    t_{sync} & \simeq  \frac{\gamma m_e c^2}{P_{sync}} \simeq \frac{9}{4} \, \frac{m_e^3 c^5}{e^4 B_s^2 \gamma} \simeq \\
    & \simeq 2.2 \, \Biggr(\frac{\epsilon}{10 \, \mathrm{keV}} \Biggr)^{-1/2} \, \Biggr(\frac{B_s}{4.5 \times 10^5 \, \mathrm{G}} \Biggr)^{-3/2} \, \mathrm{\mu s}.
\end{split}    
\end{equation}
The electron energy is $\gamma m_e c^2$ and $P_{sync} \simeq 4/3 \, \sigma_T c \gamma^2 U_B$ is the average synchrotron power per relativistic electron in a source with an isotropic pitch-angle distribution. Here $\sigma_T$ is the Thomson scattering cross section, $U_B$ is the magnetic energy density, and we used the typical energy of synchrotron photons written as $\epsilon = \hbar \omega_{sync} = 3 \hbar e B_s \gamma^2 /(2 m_e c)$.
The magnetic field intensity in the post-shock region is given by \citep{Arons_1993}
\begin{equation} \label{eq:Bs}
    B_{s} = 3 \, \Biggr( \frac{\sigma}{1+\sigma} \Biggr)^{1/2} \, \Biggr( \frac{\dot{E}}{c \, f_{p} \, r^2} \Biggr)^{1/2} \simeq 4.5 \times 10^5 \, k^{-1} \, f_{p}^{-1/2} \, \mathrm{G},
\end{equation}
where $\dot{E} = 4.43 \times 10^{34} \, \mathrm{erg \, s^{-1}}$ is the total spin-down power \citep{archibald2013longterm} and $r=k\, R_{LC}$ is the shock distance. The magnetization parameter $\sigma$ \citep{Kennel_Coroniti_1984, Bogdanov_2011} is $\gg 1$ close to the light cylinder where the wind is released \citep{Arons_2002}, and $f_{p}$ represents a geometric factor indicating the sky fraction in which the pulsar wind is emitted ($f_{p}=1$ for an isotropic pulsar wind; \citealt{Bogdanov_2011}). We assumed $1 \lesssim k \lesssim
2$, that is, the shock region where optical and X-ray pulses are produced is just beyond the light cylinder (for $k<1$ the matter of the disk would enter the light cylinder, preventing the formation of a relativistic wind from the pulsar).\\
From Eq. \eqref{eq:tsync_orig}, we found that it takes
\begin{equation}
    t_{sync}(\epsilon = 5 \, \mathrm{keV}) \simeq 3 \, \Biggr( \frac{B_s}{4.5 \times 10^5 \, \mathrm{G}} \Biggr)^{-3/2} \, \mathrm{\mu s}
\end{equation}
to emit X-ray photons. We assumed the average energy of \textit{NICER} photons to be $\epsilon \sim 5$ keV \footnote{\url{https://heasarc.gsfc.nasa.gov/docs/nicer/mission_guide}}.
On the other hand, the synchrotron timescale for optical photons, with $\epsilon \sim 1 \, \mathrm{eV}$, is $\sim 70$ times longer, such that the time lag between optical and X-ray pulsations is 
\begin{equation} \label{eq:dt_exp}
    \delta \tau \simeq 220 \, \Biggr( \frac{B_s}{4.5 \times 10^5 \, \mathrm{G}} \Biggr)^{-3/2} \, \mathrm{\mu s}.
\end{equation}
Similarly, we can predict the time lags with UV and near-infrared (NIR) pulsations from J1023+0038. Assuming an energy of $\epsilon \sim 5 \, \mathrm{eV}$ for UV photons, we would have a time lag of $\sim 122 \, \mathrm{\mu s}$ between optical and UV pulsations, and of $\sim 95 \, \mathrm{\mu s}$ between UV and X-ray pulses. For a NIR observation in the K band, the photon energy is $\epsilon \sim 0.6 \, \mathrm{eV}$: NIR pulsations would lag optical pulsations by $\sim 64 \, \mathrm{\mu s}$, UV pulsations by $\sim 186 \, \mathrm{\mu s}$, and X-ray pulsations by $\sim 281 \, \mathrm{\mu s}$.\\
When we use our parametric value of the time lag, $\delta \tau_p \sim 150 \,  \mathrm{\mu s}$, and invert Eq. \eqref{eq:dt_exp}, the intensity of the magnetic field behind the shock is 
\begin{equation} \label{eq:Bs_value}
    B_s \simeq  5.8 \times 10^5 \, \delta \tau_{p,150}^{-2/3} \, \mathrm{G},
\end{equation}
where $\delta \tau_{p,150}$ is in unit of $150 \, \mathrm{\mu s}$. For a surface magnetic field of $B_{surf} \simeq 9.6 \times 10^7$ G \citep{2012ApJ...756L..25D}, we computed the dipolar decrease at the light cylinder radius as $B_{LC} \simeq B_{surf} \, (R_{LC}/R_{NS})^{-3}$, assuming a NS radius of about $R_{NS} \simeq 10$ km. Since the magnetic field in the post-shock region is three times the intensity before the shock \citep{Arons_1993}, we found a value similar to that in Eq. \eqref{eq:Bs_value}.\\
Combining Eqs. \eqref{eq:Bs} and \eqref{eq:Bs_value}, we obtained
\begin{equation}
    f_{p} \simeq \Biggr( \frac{B_s}{4.5 \times 10^5 \, \, \mathrm{G}} \Biggr)^{-2} \, \frac{1}{k^2} \lesssim 0.6 \, \delta \tau_{p,150}^{4/3}.
\end{equation}
A geometric factor $<1$ indicates that the pulsar wind is concentrated on the equatorial plane, as also suggested by several models of the Crab pulsar \citep[see, e.g.,][]{Kirk_2006}.\\
Variations in disk truncation radius, that is, in the $k$ parameter, are to be expected as they were found in several magnetohydrodynamic simulations \citep[see, e.g.,][Fig. 2, panel (d)]{Parfrey_2017}. They clearly show that this is a highly turbulent system. This variability has been proposed to explain the disappearance of optical and X-ray pulsations during the low-intensity modes, when the inner disk boundary is pushed farther out by the pulsar wind \citep{2019A&A...629L...8C, Papitto_2019}. The disk truncation radius may vary also during the high-mode intervals, with smaller changes involved to avoid transitions between different intensity modes while causing variations in pulse amplitude \citep{Miraval_Zanon_2022} and phase.
Combining Eqs. \eqref{eq:Bs} and \eqref{eq:dt_exp}, we found that the expected time lag between optical and X-ray pulsations scales with the $k$ parameter as $\delta \tau \simeq 220 \, k^{3/2} \, f_{p}^{3/4} \, \mathrm{\mu s}$. If $k$ oscillates between 1 and 2 (i.e., the truncation radius of the disk oscillates between $\sim R_{LC}$ and $\sim 2 R_{LC}$), the time lag would vary by a factor of $\sim 2.8$. This,  as well as shock height changes, may explain some of our measured variations in the time lags (Table \ref{table:dt}). Another reason may be that some data sets do not have intervals of exact simultaneity between optical and X-ray observations. There may also be some dependence on source parameters, such as luminosity and orbital phase, which it will be necessary to investigate in future works.

J1023+0038 and the AMXP SAX J1808.4-3658 are the only two millisecond pulsars with detectable optical pulsations so far \citep{Ambrosino_2017, Papitto_2019, Ambrosino_2021}. Interestingly, optical pulsations observed from SAX J1808.4-3658 during an accretion outburst in August 2019 seem to be almost in antiphase with those in the X-rays \citep{Ambrosino_2021}, as opposed to what was observed in J1023+0038. Therefore, optical and X-ray pulsations from SAX J1808.4-3658 can hardly be explained by the same physical mechanism. An important step to confirm the origin of optical and X-ray pulsations from J1023+0038 and test the proposed model may come from the observation of other transitional millisecond pulsars in the subluminous disk state, such as XSS J12270-4859 \citep[see, e.g.,][]{2010A&A...515A..25D, 2013A&A...550A..89D, 2014MNRAS.441.1825B}, and candidates such as 3FGL J1544.6-1125 \citep[see, e.g.,][]{Bogdanov_2015, Britt_2017}, CXOU J110926.4-650224 \citep{2019A&A...622A.211C, 2021A&A...655A..52C}, PSR J0337+1715 \citep{2016MNRAS.459..427S}, and XMM J083850.4–282759 \citep{10.1093/mnras/stx1560}.

\section{Conclusions} \label{sec:conclusions}
We presented a detailed timing analysis of (quasi-)simultaneous observations in the X-rays and in the optical band of the transitional millisecond pulsar PSR J1023+0038. The analyzed data cover the time interval from May 2017 to January 2022, when the system was in a subluminous disk state. They were acquired with \textit{XMM-Newton} and \textit{NICER} X-ray satellites, and with the fast optical photometers SiFAP2 and Aqueye+. 
Our main results are summarized below.
   \begin{enumerate}
      \item Although the estimated time lags between optical and X-ray pulsations are not consistent with being modeled by a single value, the weighted averages obtained from the studies of the first and second harmonic terms are $\overline{\delta \tau}_1 =(175 \pm 22) \, \mathrm{\mu s}$ and $\overline{\delta \tau}_2 =(162 \pm 6) \, \mathrm{\mu s}$, respectively. When we focus on the second harmonic because of its higher power spectral densities, the time lag lies in a limited range of values, $\sim(0-250) \, \mathrm{\mu s}$. This is maintained over the years, supporting the hypothesis that both pulsations originate from the same region and that their emission mechanisms are intimately linked.
      \item From \textit{NICER}/Aqueye+ simultaneous observations, we found that the second harmonic of the optical pulse lags that of the X-ray pulse by $\delta \tau _2 = (254 \pm 47) \, \mathrm{\mu s}$ in February 2019, $(119 \pm 48) \, \mathrm{\mu s}$ in January 2020, and $(108 \pm 36) \, \mathrm{\mu s}$ in January 2022. These results are not affected by the systematic errors, while the previous measurement by \citet{Papitto_2019} was affected by the absolute timing uncertainty of SiFAP2 and \textit{XMM-Newton}.
      \item Our results find a convincing interpretation in the shock-driven mini pulsar nebula scenario \citep{Papitto_2019, Veledina_2019}, which suggests an origin of optical and X-ray pulses based on synchrotron radiation emitted from a shock formed where the striped pulsar wind meets the accretion disk, within a few light cylinder radii from the pulsar. The time lag is interpreted in terms of the different timescale that synchrotron X-ray and optical photons take to be emitted.
   \end{enumerate}
Within the proposed model, variations of the estimated time lags of optical pulses relative to the X-ray pulses may be due to the variability in disk truncation radius and/or to changes in the shock height. Another reason may be that some data sets lack exactly simultaneous intervals between optical and X-ray observations. There may also be some dependence on source parameters, such as luminosity and orbital phase, that must be investigated in future works.

\begin{acknowledgements}
      This work is partly based on observations made with the Italian Telescopio Nazionale Galileo (TNG), operated on the island of La Palma by the Fundación Galileo Galilei of the Istituto Nazionale di Astrofisica (INAF), which is installed in the Spanish Observatorio del Roque de los Muchachos of the Instituto de Astrofísica de Canarias, in the island of La Palma. Part of this paper is based on observations obtained with the Copernicus Telescope (Asiago, Italy) of the INAF-Osservatorio Astronomico di Padova. Some of the scientific results are obtained from observations with \textit{XMM-Newton}, which is an European Space Agency (ESA) science mission with instruments and contributions directly funded by ESA member states and NASA, and \textit{NICER}, which is a NASA mission.
      We are grateful to the \textit{NICER} directors for scheduling ToO observations in Director’s Discretionary Time, as well as to all the teams involved for their effort in scheduling the simultaneous ToO observations of PSR J1023+0038. 
      This research also made use of data and software provided by the High Energy Astrophysics Science Archive Research Center (HEASARC).
      G.I. and A.P. wish to thank A. Sanna for useful discussions.
      G.I. is supported by the AASS Ph.D. joint research program between the University of Rome “Sapienza” and the University of Rome “Tor Vergata”, with the collaboration of the National Institute of Astrophysics (INAF). F. A., G. I., A.P., L.S., D.d.M., and L.Z. acknowledge financial support from the Italian Space Agency (ASI) and National Institute for Astrophysics (INAF) under agreements ASI-INAF I/037/12/0 and ASI-INAF n.2017-14-H.0, from INAF ’Sostegno alla ricerca scientifica main streams dell’INAF’, Presidential Decree 43/2018. F. A., G. I., A.P., and L.S. also acknowledge funding from the Italian Ministry of University and Research (MUR), PRIN 2020 (prot. 2020BRP57Z). A.P. and D.d.M. acknowledge financial contribution from INAF ‘SKA/CTA projects’, Presidential Decree 70/2016. A.M.Z. is supported by PRIN-MIUR 2017 UnIAM (Unifying Isolated and Accreting Magnetars, PI S. Mereghetti). F.C.Z. is supported by a Juan de la Cierva fellowship. The work of F.C.Z and D.F.T. is also partially supported by the program Unidad de Excelencia María de Maeztu (CEX2020-001058-M). D.F.T. work has been supported by the grants PID2021-124581OB-I00 and acknowledges as well USTC and the Chinese Academy of Sciences Presidential Fellowship Initiative 2021VMA0001.
      M.T. acknowledges funding from the ERC under the EU’s Horizon 2020 research and innovation programme (consolidator grant agreement no. 101002352).
      
\end{acknowledgements}
  \bibliographystyle{aa} 
  \bibliography{bibliography.bib} 
\end{document}